Submitted to IEEE Communications Magazine for Possible Publication

# Cross-Medium Technology Transfer for RF Integrated Sensing and Communications

J. Andrew Zhang, David Plets, Chao Lu, and Andrea M. Tonello

***Abstract*— Integrated sensing and communications (ISAC) spans radiofrequency (RF), visible-light, optical-fiber, power-line, and acoustic systems, yet technology transfer across these domains remains underexplored. This article examines bidirectional knowledge and technology transfers centred on RF-ISAC. It shows how VLC motivates power-domain sensing, fiber enables differential referencing and distributed processing, and PLC inspires topology-aware monitoring and adaptive probing. Conversely, RF-ISAC contributes joint signal design, nuisance suppression, and weak-return recovery. The article highlights transferable principles, required adaptations to improve ISAC, and information potentially lost across physical media.**

## I. Introduction

Integrated sensing and communications (ISAC) has emerged as a key capability of future wireless networks. By reusing communication signals, hardware, spectrum, and/or infrastructure to sense the physical environment, radio-frequency (RF) ISAC can support localization, tracking, imaging, activity recognition, and environmental monitoring alongside conventional data transmission [1]. Nevertheless, practical RF-ISAC remains constrained by several persistent problems. In particular, bistatic and networked sensing is highly sensitive to clock asynchronism, carrier- and sampling-frequency offsets, calibration errors, and random phase variations. Monostatic sensing must recover weak reflections in the presence of strong self-interference, while multipath, static clutter, and environmental dynamics complicate the interpretation of communication channel measurements.

Related problems arise in communication systems operating over other physical media, but they are not always formulated or addressed in the same way. Visible-light communication (VLC) systems can sense people and objects from received intensity, blockage, shadows, or spatial image patterns, often without recovering coherent propagation phase [2]. Power-line communication (PLC) systems infer network topology, cable faults, impedance variations, and equipment conditions from changes in channel responses or reflectometric measurements, commonly with respect to a reference network state [3]. Optical-fibre systems detect distributed vibration, strain, temperature, and acoustic disturbances through backscattered intensity, phase, polarization, and frequency shifts, while using differential and reference processing to suppress laser and interrogator instabilities [4]. Acoustic communication and sensing systems operate under long propagation delays, severe multipath, Doppler scaling, narrow bandwidth, and rapidly varying environments, forcing joint treatment of synchronization, channel variation, and target motion [5].

These systems suggest that some unresolved RF-ISAC problems may benefit from solutions developed outside the radio domain. Technology transfer across physical media, however, cannot be justified by superficial similarity. A phase offset in RF, a laser-phase fluctuation in fibre, and a timing error in PLC may have related mathematical structures but different physical origins and reference measurements. Moreover, suppressing a nuisance parameter can also remove desired sensing information. Magnitude-only RF processing, for instance, may eliminate random carrier phase but generally cannot recover the absolute path delay. Successful knowledge transfer therefore requires identifying what information is preserved and sacrificed, and whether the source-domain technique remains physically implementable under the destination system's waveform and hardware constraints. Existing surveys primarily review various systems independently, or compare their architectures and applications. Although recent studies have begun to study RF- and optical-ISAC together [1], they have paid limited attention to identifying concrete, bidirectional technology transfers, the mechanisms enabling them, and their applicability boundaries.

This article addresses this gap by placing RF-ISAC at the centre of a bidirectional cross-medium transfer study. As summarized in Table 1, VLC motivates power-domain RF sensing through reference illumination, structured transmission, and spatial power patterns; fiber sensing informs differential coherent processing, distributed referencing, and adaptive baselines; and PLC sensing enables topology-aware multipath interpretation and communication-medium monitoring. Conversely, RF-ISAC contributes joint waveform and resource design, synchronization suppression, weak-return recovery, and networked sensing. Rather than

Prof. Zhang (Senior member, IEEE) is with University of Technology Sydney, Australia. Email: Andrew.Zhang@uts.edu.au. Prof. Plets (Senior Member, IEEE) is with imec-WAVES/Ghent University, Belgium. Email: david.plets@ugent.be. Prof. Lu (Fellow of OSA) is with Hongkong Polytechnic University. Email: chao.lu@polyu.edu.hk. Prof. Tonello (Senior member, IEEE) is with University of Klagenfurt, Austria. Email: Andrea.Tonello@aau.at. Their areas of expertise relevant to this article are RF, visible-light, optical-fiber, and power-line sensing and communications, respectively.

*Table 1 Comparison of ISAC Systems across Physical Media.*

| ISAC system | Propagation medium and signal | Principal sensing observables | Representative sensing targets | Distinctive characteristics and constraints |
|---|---|---|---|---|
| **RF-ISAC** | Unguided radio waves; commu. signals or dedicated pilots | Delay, Doppler, angle, complex CSI, received power, polarization | People, vehicles, objects, activities, and environments | Wide coverage and network integration; affected by multipath, clutter, clock asynchronism, calibration errors, and self-interference |
| **VLC-ISAC** | Intensity-modulated visible light; photodiode or camera reception | Received intensity, RGB diversity, blockage, shadows, spatial light patterns, image features | Occupancy, pose, activity, location, and object geometry | Direct detection without phase recovery; requires nonnegative signals and must satisfy illumination, dimming, field-of-view, and eye-safety constraints |
| **Fiber ISAC** | Guided optical waves and distributed backscatter | Optical intensity, phase, polarization, frequency shift, and propagation time | Vibration, strain, temperature, and infrastructure disturbances | Extremely high sensitivity and distributed spatial resolution; constrained by laser coherence, nonlinearities, and launch power |
| **PLC-ISAC** | Conducted electromagnetic signals over electrical wiring | Channel transfer function, reflection coefficient, impedance response, delay, and spectral variation | Cable faults, topology, load changes, equipment condition, and network anomalies | Communication medium is often the sensing target; affected by branching topology, impulsive noise, time-varying loads, and emission limits |
| **Acoustic ISAC** | Mechanical waves in air or water | Delay, wideband Doppler scaling, angle, amplitude, and spectral signatures | Objects, motion, environments, and events | Low propagation speed, long latency, severe multipath, limited bandwidth |

comprehensively surveying these systems, the article distils their transferable physical and signal-processing principles, clarifies the similarities and medium-specific differences governing their applicability, and shows how cross-medium transfer can expand sensing capabilities and improve ISAC performance. Acoustic and RF ISAC already share delay-Doppler-angle estimation, matched filtering, array processing, and active/passive sensing, reflecting the long-standing exchange between sonar and radar; acoustic ISAC is therefore not examined separately.

The remainder of this article is organized as follows. Sections II-IV examine VLC-inspired power-domain sensing, fiber-inspired differential and distributed sensing, and PLC-inspired topology-aware sensing for RF-ISAC, respectively. Section V explores how RF-ISAC techniques can, in turn, be adapted to other physical media. Finally, Section VI presents the conclusions and research outlook.

## II. VLC-Inspired Power-Domain RF Sensing

High-resolution RF sensing commonly uses complex channel state information (CSI), since delay, Doppler, and angle are encoded in phase variations across subcarriers, time, and antennas. In bistatic systems, however, these variations are corrupted by timing offset (TO), carrier-frequency offset (CFO), and phase offsets. Techniques addressing these offsets for RF-ISAC have been developed, see a summary in [6]. VLC-ISAC suggests an alternative: work directly with received power. Most VLC systems employ intensity modulation and direct detection. Photodiodes measure received optical power, while cameras provide spatially resolved intensity measurements. Objects modify this power through blockage, shadows, reflection, and changes in propagation geometry.

The direct RF counterpart of optical intensity is its signal power, rather than CSI power alone. In both VLC and RF systems, however, the instantaneous received power depends not only on the propagation channel but also on the transmitted communication signals. VLC sensing therefore uses temporal averaging, filtering, coded source separation, or waveform normalization to suppress data-induced fluctuations and reveal slower target-induced changes in the illumination field. RF-ISAC can similarly construct waveform-normalized power measurements. However, unlike VLC, because RF signals combine coherently from multiple propagation paths, the resulting cross-products can convert target-dependent phase changes into observable power variations.

To see this, consider a bistatic channel containing a static component, including the direct link and stationary reflections, and a dynamic component induced by a target. Self-conjugation removes common clock- and hardware-dependent phase terms. However, cross-product terms between static and dynamic paths retain information dependent on their relative phase, which is critical for retrieving phases for sensing. In particular, a strong static path acts as an implicit reference that converts target-dependent phase evolution into observable power variation. This is analogous to VLC sensing, where a target becomes detectable by perturbing a strong and otherwise stable background illumination, but with an important extension: in RF, coherent interference with the reference path also preserves relative-phase information.

Power-domain RF sensing is thus most informative when a stable reference path exists. The feasibility has been demonstrated using bistatic LTE and WiFi signals [7]. Processing per-subcarrier CSI power enables extracting delay, angle-of-arrival (AoA), and Doppler features for passive tracking without explicit phase-offset compensation or antenna-phase calibration. This motivates a specific transfer: *Treat RF power as an interference-encoded sensing observable and create or expose the required information through reference paths, structured illumination, power-field imaging and adaptive background modelling.*

### II.A. From Structured Lighting to Structured RF Illumination

VLC systems distinguish multiple luminaires by activating them sequentially or assigning separable modulation patterns. Similarly, RF transmitters, access points, UEs, or beams can be separated through time slots, subcarrier groups, codes, packet sequences, or carrier frequencies.

The receiver can then construct a power response for each transmitter-receiver-beam combination. Independently clocked transmitters do not need to be combined coherently; each provides a different bistatic illumination geometry. The observations form an "RF power cube" indexed by transmitter, antenna, beam, subcarrier, and time. Transmitters provide geometric diversity, antennas and beams provide spatial patterns, subcarrier power retains relative-delay oscillations, and temporal variation retains motion information.

This architecture is suitable for asynchronous multi-node sensing: each link first extracts power-domain features, which are then fused noncoherently or through geometry-aware tracking. Its cost is the time, frequency, code, or beam resources required to separate illuminators, motivating joint communication and sensing scheduling.

### II.B From Shadow Maps to RF Power-Field Imaging

VLC systems combine power changes across multiple links into shadow or occupancy maps rather than estimating a parameter from each link independently [8]. RF-ISAC can similarly construct "power-field images" from target-induced changes across links, antennas, beams, and frequencies.

RF shadows are more complicated than optical shadows because a target may block one path while strengthening others through reflection or diffraction. Reconstruction should therefore combine geometric partial-blockage models with calibrated or learned multipath sensitivity maps. Analogous to constructing an optical illumination map, differentiable ray-tracing tools such as Sionna RT can generate high-resolution, site-specific RF power fields from the environment geometry, material properties, antenna patterns, and deployment configuration. Measurements can then calibrate these simulated fields and compensate for modelling errors or environmental changes. Beam sweeping can further create controllable power patterns, while reconfigurable surfaces may deliberately vary the illumination field to resolve otherwise ambiguous target states. Applications include occupancy imaging, device-free localization, trajectory recovery, and intrusion mapping without coherent phase calibration.

Adaptive background modelling can support this imaging process by learning the normal power field across links, antennas, beams, and frequencies. Rather than subtracting a single reference snapshot, the system continuously updates the mean, variability, or low-dimensional subspace of target-free measurements to accommodate slow changes in transmit power, receiver gain, static multipath, and the surrounding environment. The current power field can then be normalized, differenced, or statistically whitened against this model before image reconstruction, making spatially consistent target-induced changes more visible. To avoid incorporating persistent or slowly moving targets into the background, model updates can be slowed or suspended in regions where occupancy is detected.

## III. Fiber-Inspired Differential and Distributed RF Sensing

Optical fibers can act as distributed sensors of temperature, strain, vibration, pressure, and acoustic disturbances [9, 10]. Rayleigh backscatter is commonly used for vibration and acoustic sensing, while Brillouin and Raman scattering support strain and temperature measurements. Although these physical mechanisms do not have direct RF equivalents, their sensing architectures and processing principles motivate several useful transfers to RF-ISAC.

The most immediately feasible fiber-to-RF technology transfers are differential referencing, distributed change detection, multiscale processing, and adaptive baseline management. Exploiting richer RF physical observables is more exploratory because their relationships with environmental variables may be less stable. Overall, optical-fiber ISAC demonstrates how communication infrastructure can preserve weak coherent environmental information through local referencing and organize large-scale sensing around distributed changes rather than absolute measurements.

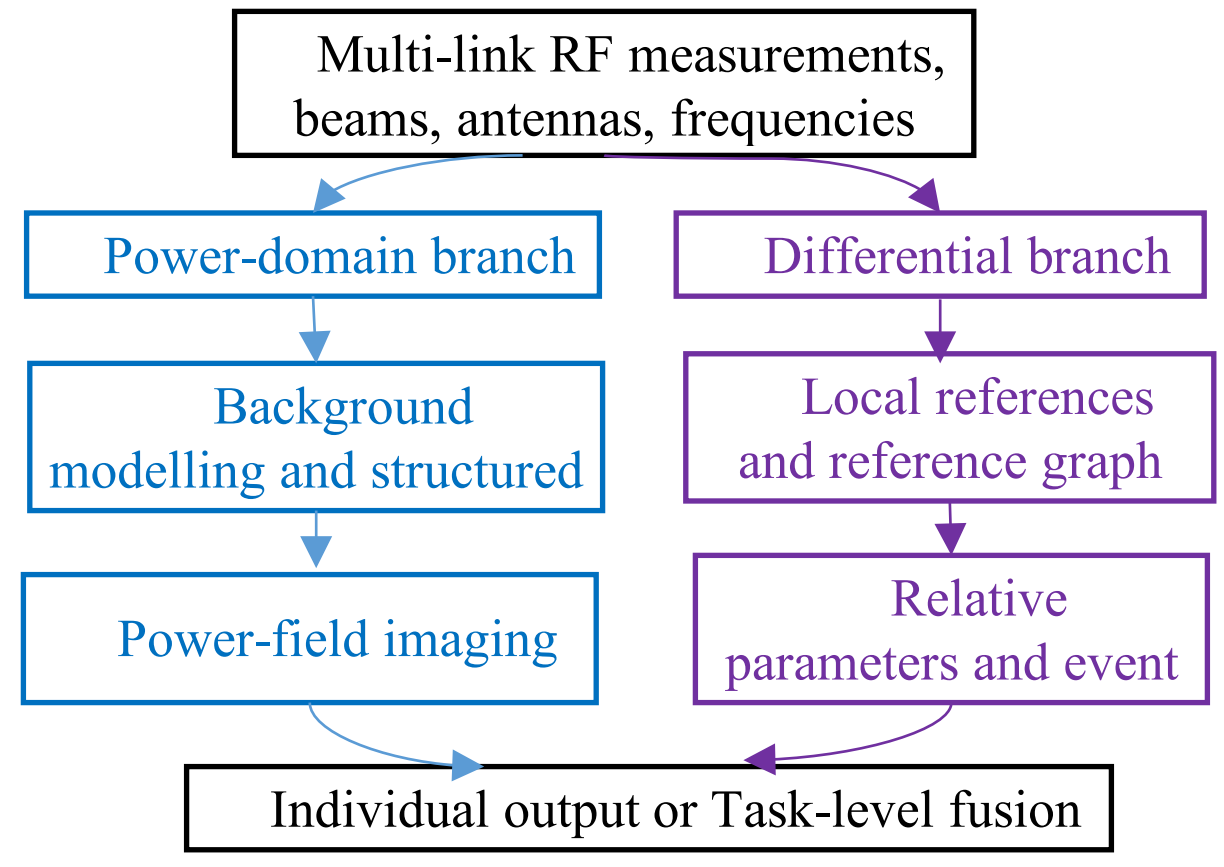

*Fig.1VLC-inspired (left) and fiber-inspired (right) realization of RF sensing.*

Fig. 1 illustrates how the VLC- and fiber-inspired techniques can be incorporated into a common RF sensing architecture. Multi-link measurements across transmitters, antennas, beams, and frequencies may be processed through either a synchronization-robust power-domain branch, motivated by VLC sensing, or a differential branch that uses local references and reference graphs, motivated by fiber sensing. The former supports power-field imaging when reliable phase is unavailable, whereas the latter preserves relative coherent information for parameter estimation and distributed event mapping. Their outputs can be individually processed or fused at the feature or task level according to the available coherence and sensing requirements. The following subsections elaborate on the differential, distributed, and multiscale processing underlying the fiber-inspired branch.

### III.A Differential Processing and Reference Graphs

Phase-sensitive fiber sensing rarely interprets the absolute phase at every position independently. Instead, it compares adjacent spatial samples, successive probing intervals, polarization channels, or stable fiber sections. Local disturbances remain visible, while phase noise and background variations shared by the paired measurements are suppressed.

RF-ISAC can similarly form ratios, or conjugate products between antennas sharing an oscillator, target-affected and stable paths, overlapping beams, or current and reference measurements of a slowly varying scene. Existing RF methods often use a dominant direct or static path to remove timing and phase offsets [6]. However, such a path may be blocked, affected by a target, or absent in non-line-of-sight sensing. Fiber sensing suggests constructing the reference from multiple locally stable measurements instead.

Motivated by local differential processing in fiber sensing and reference-path cancellation in bistatic RF sensing, these relationships can be organized as a reference graph [11]. Its vertices represent antennas, links, beams, propagation paths, or sensing cells, while an edge connects two measurements expected to share a nuisance component. Stable subgraphs can then be used to estimate common drift, whereas localized inconsistencies indicate target-induced or environmental changes. References can be weighted by their stability and excluded when contaminated. This is more robust than single-reference cancellation.

### III.B Distributed and Multiscale RF Sensing

A fiber interrogator can monitor thousands of positions along a cable. Processing is therefore organized around detecting, localizing, and classifying events over a distributed spatial domain [4]. RF networks do not form continuous linear sensors, but their transmitter–receiver links create an irregular distributed and networked sensing field [12].

Fiber sensing motivates treating links with overlapping sensitivity regions as spatially related measurements rather than independent views. A physical event should produce consistent changes across relevant links, whereas receiver faults, gain variations, or local interference may remain confined to one device. This supports an event-first, multiscale architecture [13]. The network initially monitors low-overhead power, covariance, or channel-change indicators. Nodes report event time, affected links or regions, coarse location, dominant temporal or spectral features, and uncertainty rather than forwarding complete CSI. Wider bandwidth, finer beam sweeps, longer processing intervals, or additional sensing nodes are activated only around a detected event.

The RF network can learn and support this process through spatial maps of detection probability, localization information, reference quality, synchronization quality, and expected uncertainty. These maps guide the selection of links, beams, bandwidths, and processing scales. Coarse and fine measurements may also be combined, exploiting the complementary sensitivity of narrow and wide bandwidths, short and long observation intervals, or broad and narrow beams. This reduces fronthaul and computation while avoiding continuous network-wide operation at maximum resolution.

### III.C Extending RF Sensing Observables

Fiber sensing converts effects traditionally treated as communication impairments, including phase fluctuation, polarization variation, attenuation, and scattering-frequency shifts, into environmental observables. The broader lesson for RF-ISAC is to look beyond conventional sensing parameters.

Potential RF observables include polarization changes, frequency-dependent reflection, channel dispersion, diffuse-to-specular scattering ratios, beam-dependent blockage, temporal nonstationarity, and hardware-dependent nonlinear responses. These features may reveal material properties, infrastructure degradation, or environmental conditions. RF phase-noise variations, e.g., have been exploited as sensing features in [14].

The key principle is therefore to avoid automatically removing every channel variation as an impairment. Before compensation, RF-ISAC should determine whether the variation has a stable and identifiable relationship with the environment. If so, it may constitute a useful sensing observable rather than merely a nuisance.

## IV. PLC-Inspired Topology-Aware RF Sensing

PLC uses electrical wiring simultaneously as a communication channel and as part of the monitored infrastructure. The channel response depends on cable lengths, branches, terminations, connected loads, and impedance discontinuities. A cable fault, equipment connection, degraded joint, or other anomaly changes the input impedance, reflection coefficient, or end-to-end channel transfer function. PLC modems can therefore infer the condition and topology of the network from signals already used for communication.

Unlike RF-ISAC, which normally senses external objects, PLC sensing frequently treats the communication medium itself as the sensing target. This distinction motivates several useful knowledge transfers to RF-ISAC: communication-channel diagnostics as sensing, topology-aware interpretation of multipath, reference-response anomaly localization, and task-oriented selection of sensing algorithms. The transfer is nontrivial in highly dynamic RF environments, where normal human activity and moving objects continually modify the channel. Reference models must distinguish temporary variations from persistent structural changes. RF topology is also less constrained than a wired branching network, making path-to-structure association more ambiguous.

Nevertheless, PLC-ISAC provides a clear broader lesson: *The communication channel is not merely a means of observing external targets; its structure, stability, and degradation can themselves constitute the sensing objective.* Rather than interpreting CSI variations primarily as signatures of moving targets, this perspective seeks persistent changes in the propagation topology, infrastructure, and surrounding environment. It can therefore extend RF-ISAC from target detection and tracking toward continuous monitoring of communication infrastructure and its surrounding environment. As illustrated in Fig. 1, a topology-aware RF sensing system first detects and localizes deviations from a learned channel-structure model, then adaptively schedules targeted measurements to confirm and classify the change. The following subsections examine these two stages.

### IV.A Topology-Aware Channel Monitoring and Anomaly Localization

A key PLC insight is that multipath carries structural information. Branches and impedance mismatches produce reflections whose delays and amplitudes reveal cable lengths, connection points, loads, and network topology. A new, displaced, or weakened reflection may therefore indicate a fault or structural change [3]. Similarly, stable RF paths are associated with walls, doors, machinery, and other reflecting structures. Rather than treating them only as clutter, RF-ISAC can organize these paths into a "multipath topology graph" describing persistent propagation structures or virtual anchors.

The current channel or topology response can then be compared with a reference model of normal conditions. This model should capture normal variability rather than represent a single fixed snapshot. After subtraction or statistical whitening, significant residuals can be detected and characterized through: path appearance or disappearance; shifts in delay or angle; changes in reflection strength; and altered connectivity among propagation regions. These changes can be localized using delay, angle, beam, or multi link consistency and classified from their spatial, spectral, and temporal signatures. Combining monostatic and bistatic observations can further improve reliability: monostatic measurements offer accurate local delay information, while bistatic links provide complementary geometry and broader coverage. Consistency across these modes can help distinguish genuine environmental changes from transceiver drift.

The broader lesson is that RF-ISAC need not always reconstruct target trajectories. By treating stable multipath as a structural baseline, communication-channel measurements already collected for link management can support continuous monitoring of infrastructure and its surrounding environment, including non-line-of-sight settings where direct target paths are weak or unavailable.

### IV.B Topology-Aware Adaptive Probing

PLC sensing does more than detecting changes in a fixed channel response. Because fault signatures depend on cable branches, termination impedances, loads, and probing frequency, the inferred network topology can guide the selection of subsequent measurements. A modem may probe frequency bands in which a suspected discontinuity produces a distinctive reflection, compare observations from different network locations, or combine reflectometric and end-to-end responses to reduce localization ambiguity.

RF-ISAC can inherit this principle as topology-conditioned measurement selection. Given a multipath-topology graph or environmental radio map, the system can identify the links, beams, subcarriers, or sensing modes that best distinguish among candidate explanations for an observed anomaly. For example, it may activate paths interacting with a monitored wall or machine, combine monostatic delay measurements with bistatic spatial diversity, or suppress paths dominated by unstable clutter.

This criterion differs from resource allocation based primarily on received signal strength or achievable resolution. The strongest link is not necessarily the most informative: a weaker path may traverse the affected region or respond more distinctively to a particular structural change. Measurement value should therefore reflect the predicted variation in path delay, gain, angle, or connectivity under competing environmental hypotheses.

The resulting process is a closed sensing loop, as illustrated in Fig. 2: channel monitoring identifies an unexpected change;

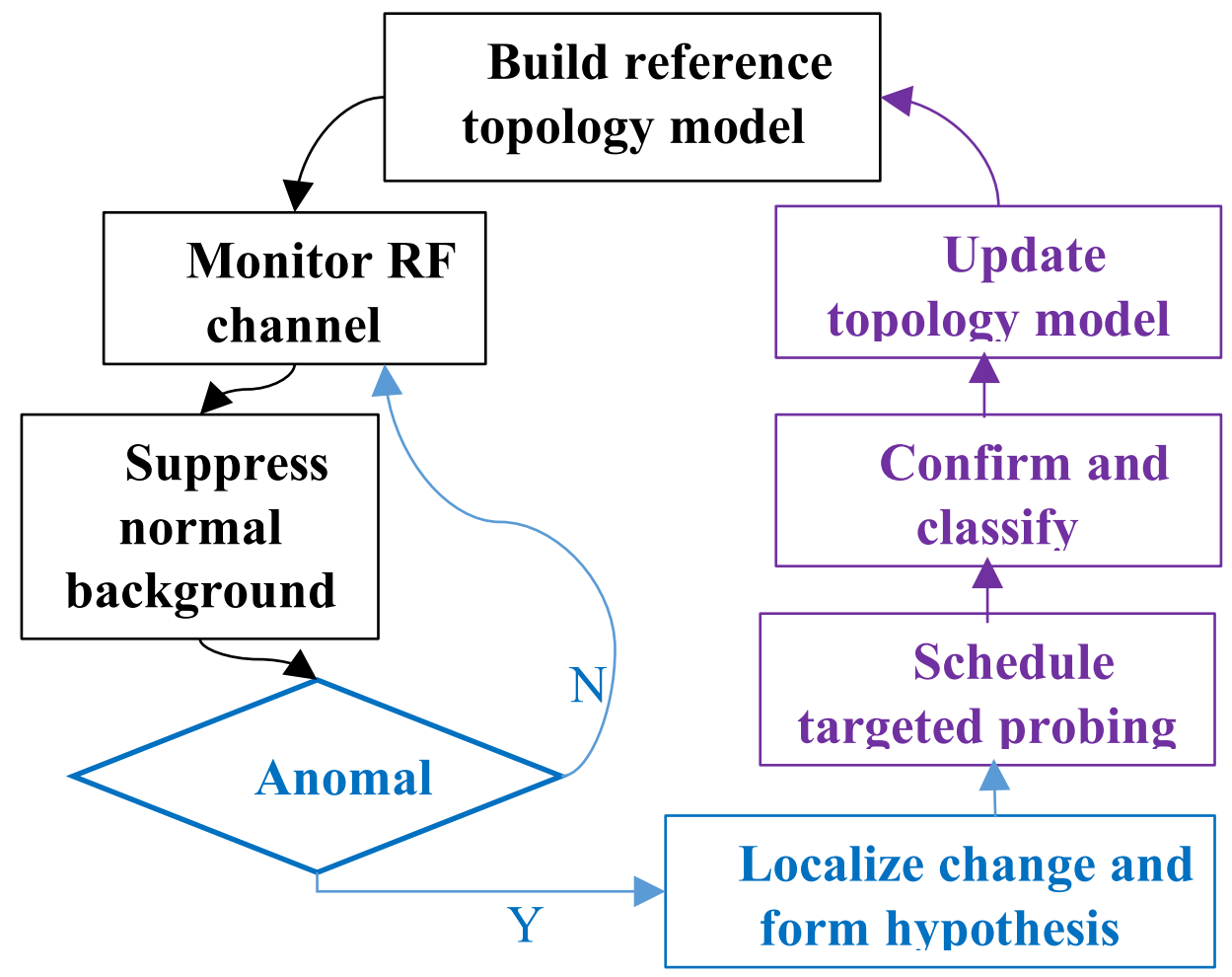

*Fig. 2. PLC-inspired Topology-aware RF channel monitoring and adaptive probing.*

the topology model associates it with candidate paths and regions; targeted probing distinguishes among the candidate causes; and confirmed observations update the structural model. The distinctive PLC-to-RF lesson is thus to use propagation structure not only to interpret anomalies, but also to decide what to measure next. This can improve localization and classification in bandwidth-limited, non-line-of-sight, and multi-link deployments without repeating the general multiscale activation architecture of Section III.B.

## V. RF-ISAC FOR OTHER PHYSICAL MEDIA

Although Sections II-IV present knowledge diffusion according to their source systems, Table 2 reorganizes them by the underlying measurement and processing principles. This cross-system view reveals five complementary approaches for RF-ISAC: incoherent power sensing, reference construction, distributed processing, topology-aware monitoring, and adaptive probing. Some methods originate primarily from one medium, whereas others emerge from related practices across several systems.

Transfer in the opposite direction is also valuable, via RF-ISAC's comparatively mature methods for joint waveform design, synchronization suppression, and weak-return recovery, as detailed next.

### V.A Joint Waveform and Resource Design

RF-ISAC commonly balances communication rate, sensing accuracy, detection performance, energy, and interference through joint waveform, beamforming, and resource optimization. The same multi-objective framework can be applied to other media.

In VLC, modulation must satisfy nonnegativity, illumination, dimming, eye-safety, LED-bandwidth, and field-of-view constraints. Joint design could optimize lighting distribution, communication symbols, modulation depth, luminaire activation, sensing diversity, and target separation simultaneously. Structured illumination could be selected according to both user data demand and the expected information it provides about objects.

In PLC, subcarrier power, pilots, and probing intervals could be optimized jointly for data delivery and sensitivity to faults or topology changes. The design must account for emission masks, frequency-selective attenuation, impulsive noise, and time-varying loads.

Fiber systems can jointly allocate wavelength, launch power, polarization, probing bandwidth, and pulse repetition. They can also jointly allocate spatial channels for spatial division multiplexed multicore and few mode fiber systems. A demonstrated system used the same linear-frequency-modulated optical carrier for PAM4 transmission and distributed vibration sensing, showing that joint design can sometimes improve communication as well as sensing [10].

### V.B Synchronization and Common-Nuisance Suppression

Bistatic RF-ISAC has exploited various techniques for suppressing asynchronous offsets for coherent sensing [6]. These methods can be used in another ISAC domain when the destination system provides multiple measurements sharing a common nuisance.

*Table 2 Cross-Medium Principles Transferable to RF-ISAC.*

| Transfer topic | Source-system insight | Realization in RF-ISAC | Principal value and applicability condition |
|---|---|---|---|
| **Power-domain sensing** | VLC infers targets from intensity perturbations, shadows, and structured illumination without recovering propagation phase. | Use CSI power or RSSI; exploit interference between stable and target-induced paths; form power fields across links, antennas, beams, and frequency. | Avoids explicit phase synchronization and enables low-cost sensing; requires stable reference paths. |
| **Reference construction, differential processing** | VLC uses background illumination; fiber uses local backscatter references; PLC compares current responses with a normal network state. | Construct references across antennas, paths, time, frequency, or spatial regions; apply ratios, conjugate products, differencing, or adaptive baselines. | Suppresses common hardware and environmental disturbances while retaining weak target changes. |
| **Distributed processing and networked sensing** | Fiber localizes events along distributed sensing paths; VLC combines multiple luminaires and receivers; PLC uses multiple modems or probing locations. | Perform local feature extraction followed by noncoherent or uncertainty-aware multi-link fusion. | Benefits depend on controlling reporting overhead and correlated information. |
| **Topology-aware channel monitoring** | PLC interprets reflections and channel-transfer changes through cable branches, network connectivity, and impedance discontinuities. | Represent stable RF multipath as a topology graph of paths, reflectors, virtual anchors, or propagation regions; detect changes in path existence, delay, angle, gain, or connectivity. | Turns channel into sensing targets; particularly useful for persistent anomaly and non-line-of-sight monitoring. |
| **Structured and adaptive probing** | VLC controls illumination patterns; PLC selects frequencies and probing locations via fault sensitivity; fiber varies spatial and temporal resolution. | Select transmitters, beams, subcarriers, bandwidth, and coherent-processing levels according to predicted event sensitivity rather than SNR alone. | Reduces sensing overhead and resolves ambiguous events through targeted follow-up measurements. |

For end-to-end PLC sensing, independent modem clocks can introduce phase slope, carrier offset, and sampling drift that are confused with propagation or fault-induced changes. Reference links, reciprocal measurements, and joint offset–anomaly estimation are therefore directly relevant.

In coherent fiber systems, RF oscillator offsets correspond approximately to laser-frequency offset, local-oscillator phase noise, and pulse-to-pulse phase variation. Polarization channels, wavelengths, spatial optical fiber channel, stable fiber sections, or reference reflectors can provide the multiple observations required for common-mode cancellation. The transfer is most valuable for multi-interrogator fiber systems; a conventional co-located interrogator already establishes coherence by construction.

VLC intensity sensing is less sensitive to carrier phase, but asynchronous luminaires and cameras still experience frame, symbol, sampling, and rolling-shutter offsets. RF reference-signal and joint synchronization-sensing methods may assist high-rate modulated-light sensing when intensity patterns alone are insufficient.

### V.C Weak-Return Recovery and Full-Duplex Processing

Monostatic RF-ISAC must recover weak echoes in the presence of a much stronger transmitted signal. Its combination of passive isolation, analogue cancellation, digital reconstruction, and residual-interference suppression provides a transferable layered architecture [15].

PLC reflectometry may need to detect weak fault reflections while communication signals and strong impedance discontinuities remain present. Communication-data-aided reconstruction can remove the known transmitted component before reflectometric processing. Fiber backscatter is extremely weak relative to forward transmission. Native optical circulators, filtering, wavelength separation, and coherent detection remain essential, but RF-style adaptive digital cancellation may suppress residual leakage using known communication data. In VLC, optical isolation and field-of-view control can similarly be complemented by subtracting a calibrated target-free illumination baseline to expose target-induced power changes and, for modulated sources, cancelling known or decoded communication components.

## VI. Conclusions and Outlook

This article examined bidirectional cross-medium technology transfer with RF-ISAC as the central focus. VLC shows that power, structured illumination, background subtraction, and spatial patterns can enable sensing without coherent phase; in RF, static–dynamic path interference allows CSI power and even RSSI to retain useful target cues. Fiber sensing contributes differential measurements, local references, adaptive baselines, and multiscale distributed processing, while PLC treats the communication medium and topology-induced multipath as sensing targets. Conversely, RF-ISAC offers joint waveform/resource design, synchronization suppression, and weak-return recovery. Transfer should focus on principles subject to each medium's physical and hardware constraints.

Several research directions deserve priority. Firstly, RF sensing should adapt its coherence level to the task, synchronization quality, and available reference paths, switching among power-domain, differential, and fully coherent processing. Secondly, reference graphs and environmental digital twins should be developed for long-term network sensing under changing hardware and environments. Third, future transfer studies should quantify not only sensing accuracy but also the information lost through nuisance suppression, normalization, aggregation, or noncoherent processing.

Cross-medium technology transfer is valuable because structurally equivalent sensing problems can sometimes share solutions. Recognizing these equivalences and limits can open new design directions for RF-ISAC while accelerating sensing integration across communication infrastructures.